\documentclass[twocolumn,showpacs,preprintnumbers,amsmath,eqsecnum,amssymb]{revtex4}
\begin{document}
\preprint{IMAFF-RCA-04-11}
\title{Dark energy and supermassive black holes}
\author{Pedro F. Gonz\'{a}lez-D\'{\i}az}
\affiliation{Colina de los Chopos, Centro de F\'{\i}sica ``Miguel A.
Catal\'{a}n'', Instituto de Matem\'{a}ticas y F\'{\i}sica Fundamental,\\
Consejo Superior de Investigaciones Cient\'{\i}ficas, Serrano 121,
28006 Madrid (SPAIN).}
\date{\today}
\begin{abstract}
This paper deals with a cosmological model in which the universe
is filled with tachyon dark energy in order to describe current
and future accelerating expansion. We obtain that the simplest
condition for the regime of phantom energy to occur in this
scenario is that the scalar field be Wick rotated to imaginary
values which correspond to an axionic field classically. By
introducing analytical expressions for the scale factor or the
Hubble parameter that satisfy all constraint equations of the used
models we show that such models describe universes which may
develop a big rip singularity in the finite future. It is argued
that, contrary to a recent claim, the entropy for a universe
filled with dark energy is definite positive even on the phantom
regime where the universe would instead acquire a negative
temperature. It is also seen that, whichever the fate of the
tachyonic accelerating universe, it will be stable to any
fluctuations of the scalar field, and that since the considered
models have all an imaginary sound speed, any overdense regions
will undergo an accelerate collapse leading rapidly to formation
of giant black holes. Finally the conjecture is advanced that
these black holes may be the supermassive black holes that most
galaxies harbor at their center.
\end{abstract}


\maketitle

\section{Introduction}

The discovery that high redshift supernovae of type Ia are less
bright than expected [1,2] has opened a really new cosmological
scenario with far-reaching implications which may even reach the
standard model of particle physics and the very nature of general
relativity itself. The straightforward interpretation of this
result is that the universe is currently undergoing a period of
accelerated expansion, similar to primordial inflation, but with a
rather uncertain future. A most conservative approach to explain
the result while keeping general relativity essentially untouched
is invoking the inclusion of a now dominating extra component with
negative pressure, usually known as dark energy [3], in the
universe. There are several candidates for dark energy whose
properties are being probed by cosmological observations,
including mainly a pure cosmological constant, quintessence [4] or
k-essence [5] scalar fields (which may [6] or may not be tracked
[7]), and some unified dark matter scenarios among which two main
plausible models have been suggested: the generalized Chaplygin
gas [8], which is described by a highly non conventional equation
of state, and the cosmic tachyon theory based on non canonical
kinetic energy [9].

On the other hand, the increasing dropping of detailed analysis
leading to quite an ample observationally acceptable parameter
space beyond the cosmological-constant barrier [10] is opening the
really intriguing possibility that the universe is currently
dominated by what is dubbed as phantom energy [11]. Phantom energy
has rather weird properties which include [12]: an energy density
increasing with time, naively unphysical superluminal speed of
sound, violation of the dominant energy condition which would
eventually allow existence of inflating wormholes and ringholes
[13], and ultimately emergence of a doomsday singularity in the
finite future which is known as the {\it big rip} [14]. The regime
for phantom energy takes place for state equation parameters
$\omega=p/\rho<-1$ and has been shown to occur in all current
dark-energy models. However, whereas the big rip singularity is
allowed to happen in quintessence [14] and k-essence [15] models,
it is no longer present in models based on generalized
Chaplygin-gas equations of state [8,16] having the form
$P=-A/\rho^n$, with $A$ and $n$ being constants. No discussion has
been so far made nevertheless on the occurrence of phantom energy
and big rip in the other major contender model for unified dark
energy: the tachyon matter scenario of Padmanabhan {\it et al.}
and others [9,17] or its sub-quantum generalization [18]. Abramo
and Finelli have in fact used [19] a Born-Infeld Lagrangian with a
power-law potential and recovered a nice dark-energy behaviour,
but did not consider the negative kinetic terms which appear to
characterize phantom energy.

While defining a phantom energy regime in such scenarios appears
to be rather straightforward, it is quite more difficult to obtain
an associated expression for the scale factor of the accelerating
universe which allows us to see whether or not a big rip
singularity may occur. Under the assumption of a constant
parameter for the equation of state, we derive in this paper a
rather general solution for the scale factor of a universe
dominated by tachyon matter, and show that, quite similarly to as
it happens in current quintessence models [2,3,6,7], that solution
may predict both an ever expanding accelerating behaviour and the
occurrence of a big rip singularity, theoretically covering
therefore all observationally allowed possibilities other than a
future big crunch.

Unified dark matter or energy models are very interesting [8,9] in
that they are able to simultaneously describe the properties of
both dark matter and some form of dark energy (often a
cosmological constant) as two opposite limiting situations (at
high and low densities respectively) arising from just one theory.
The generalized cosmic Chaplygin models have been however claimed
to be observationally inconsistent and actually unstable to
short-scale fluctuations [20]. This conclusion was extended to
also encompass any unified dark matter or energy models, including
the cosmic tachyon scenario [20]. However, that conclusion is true
only when both baryons and dark matter are assumed to vanish in
the Chaplygin model. If a given, finite fraction of either is
introduced in this model then the oscillations may turn out to
become very small and do not necessarily spoil the spectra
[21,22].

In the present paper we shall use our general solution for the
scale factor in a universe filled with tachyon dark energy in
order to analyze in some detail the nature and evolution of the
fluctuations. We obtain that the unified tachyon theory is stable
to any fluctuations and that these may contribute the structure in
the universe in a rather decisive way. Actually we conjecture that
the necessary accelerate collapse of the formed overdense patches
may lead to the formation of the kind of giant black holes which
are now believed to occur at the center of most galaxies [23].

The paper can be outlined as follows. In Sec. II we discuss a
rather general solution for the scale factor which satisfies all
requirements and constraints imposed by tachyon theory. The
phantom regime for such a solution is then investigated in this
section where it is also seen that it shares all funny properties
of quintessential phantom energy, including the new characteristic
of having a negative temperature, even though its entropy keeps
being definite positive for any value of the equation of state
parameter. The evolution of fluctuations with small wavelengths is
considered in some detail in Sec. III, using both a synchronous
perturbation formalism and the relativistic analog of the
Newtonian first order perturbation equation. Sec. IV contains a
discussion on the gravitational spherical collapse of formed
overdensity regions which is further accelerated by pressure
negativeness and the very nature of the tachyonic stuff. The
conjecture that the giant black holes formed as a result of that
collapse could actually be those black holes which most galaxies
harbor at their central bulge is advanced in Sec. V. Results are
summarized and briefly discussed in Sec. VI.

\section{Tachyon model for dark energy}

For the most favored cosmological spatially flat scenario, the
Friedmann equations read
\begin{equation}
H^2\equiv\left(\frac{\dot{a}}{a}\right)^2=\frac{8\pi G\rho}{3}
,\;\; \frac{\ddot{a}}{a}=-\frac{4\pi G(\rho+3P)}{3} ,
\end{equation}
where $\rho=\rho_{NR}+\rho_{R}+\rho_{\phi}$ is the energy density
for, respectively, non-relativistic, relativistic and tachyon
matter, and $P$ is the corresponding pressure. We shall restrict
ourselves to consider a description of the current cosmic
situation where it is assumed that the tachyon component largely
dominates and therefore we shall disregard in what follows the
non-relativistic and relativistic components of the matter density
and pressure. For the tachyon field $\phi$ we have [17]
\begin{equation}
\rho=\frac{V(\phi)}{\sqrt{1-\dot{\phi}^2}} ,\;\;
P=-V(\phi){\sqrt{1-\dot{\phi}^2}} ,
\end{equation}
in which $V(\phi)$ is the tachyon potential energy. Assuming an
equation of state $P=\omega\rho$ for the tachyon matter, we then
deduce that
\begin{equation}
\omega=\dot{\phi}^2-1 .
\end{equation}
Finally, the equation of motion for $\phi$ is
\begin{equation}
\ddot{\phi}+\left(1-\dot{\phi}^2\right)\left[3H\dot{\phi}
+\frac{1}{V(\phi)}\frac{dV(\phi)}{d\phi}\right] = 0.
\end{equation}

\subsection{$\omega>-1$}

We shall show next that there exists a general solution for the
scale factor $a(t)$ in this tachyon-field scenario which has
exactly the same dependence on time as that has been obtained in
the general solution for a pure quintessence scalar field, and
hence we also show that such a scenario admit the existence of a
tachyon phantom field which leads to a singularity in finite time.
In fact, a recipe has been provided by Padmanabhan himself [17]
according to which, given the explicit form for the scale factor
$a(t)$, a complete specification of the full $\phi$-field theory
can be achieved by using the following relations:
\begin{equation}
\frac{\dot{\rho}}{\rho}=2\frac{\dot{H}}{H}
\end{equation}
\begin{equation}
\dot{\phi}=\left(-\frac{2}{3}\frac{\dot{H}}{H^2}\right)^{1/2}
\end{equation}
\begin{equation}
V=\frac{3H^2}{8\pi
G}\left(1+\frac{2}{3}\frac{\dot{H}}{H^2}\right)^{1/2} .
\end{equation}
Our task then is to choose a general expression for $a(t)$ which
simultaneously satisfies relations (2.5), (2.6) and (2.7),
together with the Friedmann equations (2.1) and the equation of
motion for the tachyon field (2.4), which be able to match the
accelerating behaviour of the current universe and implies a
physically reasonable and suitably motivated field potential. If
we assume a linear time-dependence of the tachyon field $\phi$ and
hence constancy of parameter $\omega$, then it is not difficult to
check that a general form of such an expression for $a(t)$ can be
written as
\begin{equation}
a(t)=\left[a_0^{3\left(1+\omega\right)/2}+ \frac{3}{2}\left(1+
\omega\right)t\right]^{2/\left[3\left(1+ \omega\right)\right]} ,
\end{equation}
where $a_0$ is the initial value of the scale factor at the onset
of tachyon dark energy domination. We note that this solution
describes an accelerating universe in the interval
$-1/3>\omega>-1$. At the extreme point $\omega=-1/3$, $a(t)$
describes a universe whose size increases just as $t$, such as it
should be expected. It is worth realizing that by simply trivially
re-scaling the time parameter, solution (2.8) turns out to be
nothing but the scale factor that represents the most general
solution for the case of a quintessence scalar field for a
constant equation of state [13]. On the other hand, for a scale
factor (2.8) the tachyon field and potential are given by
\begin{equation}
\phi=\phi_0+\sqrt{1+\omega}\;t
\end{equation}
\begin{equation}
V(\phi)=\frac{3\sqrt{-\omega}}{8\pi
G\left[a_0^{3(1+\omega)/2}+\frac{3}{2}\sqrt{1+
\omega}(\phi-\phi_0)\right]^2} .
\end{equation}
We note that as $\phi\rightarrow\infty$ this potential reasonably
vanishes after taking the form already considered by Padmanabhan
and others [17]. As $\phi\rightarrow\phi_0$ at $t=0$, $V(\phi)$
tends to a finite constant value, so clearly separating from the
unphysical behaviour of the potential considered by Padmanabhan
and compatible with what can be supported by string theories [17].
We regard therefore potential (2.10) as being physically
reasonable. Finally, using Eqs. (2.2), (2.3) and (2.8)-(2.9), we
obtain for the speed of sound
\begin{equation}
c_s^2=\frac{\dot{P}}{\dot{\rho}}=\omega .
\end{equation}
This result is in contradiction with the value $c_s^2=-\omega$
considered in Refs. [20,21]. The reason for that discrepancy comes
from the definitions used in this paper of the Lagrangian [17],
$L=P=-V(\phi)\sqrt{1-\dot{\phi}^2}$, and the energy density given
in Eq. (2.2).

For the accelerating-expansion regime, we see thus that the speed
of sound becomes imaginary, a case which could imply an
accelerating collapse of the tachyon stuff that can still be
circumvented however [24]. Eq. (2.11) is compatible with all the
above requirements, provided we assume a linear time-dependence
for the field $\phi$. On the other hand, since $\omega<0$, this
equation may be also compatible with a wave equation for tachyon
fluctuations where there is factor $-\omega$ in front of the
Lagrangian, such as the one used by Frolov, Kofman and Starobinsky
[20].

\subsection{$\omega<-1$}

The phantom energy regime will be characterized by values of the
state equation parameter such that $\omega<-1$ and a consequent
violation of the dominant energy condition, i.e.
\begin{equation}
P+\rho=\frac{V(\phi)\dot{\phi}^2}{\sqrt{1- \dot{\phi}^2}}<0 .
\end{equation}
Such a regime can be obtained by simply Wick rotating the tachyon
field so that $\phi\rightarrow i\Phi$, with which the field $\Phi$
can be viewed as an axion tachyon field [25], as the scale factor
$a(t)$ and the field potential $V(\Phi)$ keep being positive and
given respectively by
\begin{equation}
a(t)=\left[a_0^{-3\left(|\omega|-1\right)/2}-
\frac{3}{2}\left(|\omega|-1\right)t\right]^{-2/\left[3\left(|\omega|
-1 \right)\right]} ,
\end{equation}
which accounts for a big rip singularity at finite future time
\begin{equation}
t_* =\frac{2}{3(|\omega|-1)a_0^{3(|\omega|-1)/2}} ,
\end{equation}
and
\begin{equation}
V(\Phi)=\frac{3\sqrt{|\omega|}}{8\pi
G\left[a_0^{-3(|\omega|-1)/2}-\frac{3}{2}\sqrt{|\omega|-
1}(\Phi-\Phi_0)\right]^2} ,
\end{equation}
with $\Phi_0\rightarrow -i\phi_0$. We note that both this
potential and the phantom tachyon energy density,
\begin{equation}
\rho_{\Phi}=\frac{3}{8\pi G\left[a_0^{-3(|\omega|-1)/2}
-\frac{3}{2}(|\omega|-1)t\right]^2} ,
\end{equation}
(which has been obtained by using Eqs. (2.2), (2.3) and (2.15))
increase with time up to blowing up at the singularity at
$t=t_{*}$, to steadily decrease toward zero thereafter, in a
regime where the universe contacts to finally vanish as
$t\rightarrow\infty$. Thus, the tachyon model for dark energy
contains a regime for phantom energy which preserves all the weird
properties shown by this in current quintessence and k-essence
scenarios; i.e. superluminal imaginary speed of sound, increasing
energy density, violation of dominant energy condition and a big
rip singularity followed by a contracting phase.

It has been quite recently pointed out [26] that the entropy for
phantom energy is definite negative, and that therefore a cosmic
regime with phantom energy should be excluded. However, there
exists a rather general argument, valid for all phantom models
studied so far including the one considered in this paper, which
appears to prevent these conclusions. In fact, for a general FRW
flat universe filled with a dark energy satisfying the equation of
state $p=\omega\rho$ (with $\omega$=const.) where the first law of
thermodynamics holds and the entropy per comoving volume stays
constant [27], the temperature of the universe is given by
\begin{equation}
{\rm T}= \kappa(1+\omega)a^{-3\omega} ,
\end{equation}
where $\kappa$ is a positive constant. From Eq. (2.17) it is
already seen that ${\rm T}$ is negative for $\omega<-1$. On the
other hand, it is well known that by integrating the cosmic law of
energy conservation, $\dot{\rho}+3\rho(1+\omega)H=0$, we have
\begin{equation}
\rho=\rho_0 a^{-3(1+\omega)} .
\end{equation}
Now, from Eqs. (2.17) and (2.18) one can readily derive the
following generalized Stefan-Boltzmann law
\begin{equation}
\rho=\rho_0\left(\frac{{\rm T}}{\kappa(1+
\omega)}\right)^{\frac{1+\omega}{\omega}} .
\end{equation}
Note that for $\omega=1/3$ Eq. (2.19) consistently reduces to the
usual law for radiation, and that for $0>\omega>-1$ the dark
energy density decreases with the temperature, which is already a
rather weird behaviour. However, for the regime where $\omega<-1$,
in order to preserve $\rho$ positive, we must necessarily take
${\rm T}<0$, and therefore $\rho$ will increase with $|{\rm T}|$.
Since on the phantom regime $\rho$ increases with the scale factor
$a(t)$ it also follows that $|{\rm T}|$ increases as the universe
expands on that regime. Finally, a general expression for the dark
energy entropy can also be obtained [26] which reads
\begin{equation}
S=C_0\left(\frac{{\rm T}}{1+\omega}\right)^{1/\omega}V ,
\end{equation}
where $C_0$ is a positive constant and $V$ is the volume of the
considered portion within the dark energy fluid. It follows that,
contrary to the claim in Ref. [26], the entropy of a dark-energy
universe is {\it always} positive, even on the phantom regime.
Actually, by inserting Eq. (2.17) into Eq. (2.20) one attains that
$S$=Const. when we take $V$ as the volume of the entire universe.

It is nonetheless the temperature which becomes negative instead
of entropy for $\omega<-1$. Even though it is not very common in
physics and therefore can be listed as just another more weird
property of the phantom scenario, a negative temperature is not
unphysical or meaningless. Systems with negative temperature have
already been observed in the laboratory and interpreted
theoretically. In the case of phantom energy it means that the
entropy of a phantom universe monotonically decreased if one would
be able to add energy to that universe. Hence, a $\omega<-1$
universe would always be "hotter" that any $\omega>-1$ universe,
and if two copies of the universe were taken, one with positive
and other with negative temperature, and put them in thermal
contact, then heat would always flow from the negative-temperature
universe into the positive-energy universe. It could yet be argued
that negative temperature is a quantum-statistical mechanics
phenomenon and therefore cannot be invoked in the classical realm.
However, a negative temperature given by Eq. (2.17) when
$\omega<-1$ [27] can still be heuristically interpreted along a
way analogous to how e.g. black hole temperature can be
interpreted (and derived) without using any quantum-statistical
mechanics arguments; that is by simply Wick rotating time,
$t\rightarrow i\tau$, and checking that in the resulting Euclidean
framework $\tau$ is periodic with a period which precisely is the
inverse of the Hawking temperature [28]. Thus, the Euclideanized
black hole turns out to be somehow "quantized". Similarly in the
present case, the phantom regime can be obtained by simply Wick
rotating the classical scalar field, $\phi\rightarrow i\Phi$,
which, by Eq. (2.9), is equivalent to rotating time so that
$t\rightarrow i\tau$, too. It is in this sense that the phantom
fields are also "quantized" and that the emergence of a negative
temperature in the phantom regime becomes consistent.

\section{Renaissance of the unified tachyon dark matter model}

The biggest problem which has been claimed to be confronted by
tachyon theory of dark energy refers to its property of being a
unified dark matter model with a nonzero speed of sound. In fact
such a kind of model has been ruled out as causing violently
unphysical blowup in the matter power spectrum [20]. Actually, the
precise way in which the unified dark matter models separates from
what is observed has been only carried out for the case of the
so-called cosmic generalized Chaplygin gas [20] whenever baryons
and dark matter are assumed to be absent. Rather general arguments
were then raised [20] to generalize the conclusion to any unified
dark energy model whose equation of state allows for nonzero
values of the sound speed. In what follows we shall show that,
even though baryons and dark matter are not explicitly introduced,
such a generalization is no longer valid, at least for unified
models of tachyon dark matter and energy. We will in fact see that
perturbations with wavelength below the Jeans scale do not
increase along the available time interval for the imaginary
values taken by the sound speed for the tachyon field for any
$\omega$ and a scale factor as given by Eq. (2.8). The stress
tensor for the tachyon scalar field can be written in the perfect
fluid form [17]
\begin{equation}
T_k^i =(p+\rho)u^i u_k -p\delta_k^i ,
\end{equation}
where the pressure and energy density are given in Eq. (2.2) and
\begin{equation}
u_k=\frac{\partial_k\phi}{\dot{\phi}} .
\end{equation}
This stress tensor can be now split into a pressureless, dark
matter component and a dark energy component [17]:
$\rho=\rho_v+\rho_{\rm DM}$ and $p=p_v+p_{\rm DM}$. Now, the
vacuum component can correspond to either a cosmological constant
if we choose [17]
\begin{equation}
\rho_{\rm DM}=\frac{V(\phi)\dot{\phi}^2}{\sqrt{1-\dot{\phi}^2}}
,\;\; p_{\rm DM}=0
\end{equation}
\begin{equation}
\rho_v=V(\phi)\sqrt{1-\dot{\phi}^2} ,\;\; p_v=-\rho_v
\end{equation}
(for which $\omega_v=-1$), or to a slowly-varying quintessential
scalar field when we choose
\begin{equation}
\rho_{\rm DM}=V(\phi)\sqrt{1-\dot{\phi}^2} ,\;\; p_{\rm DM}=0
\end{equation}
\begin{equation}
\rho_v=\frac{V(\phi)\dot{\phi}^2}{\sqrt{1-\dot{\phi}^2}} ,\;\;
p_v=-V(\phi)\sqrt{1-\dot{\phi}^2}.
\end{equation}
In this latter case we have
\begin{equation}
\omega_v=1-\dot{\phi}^{-2} ,
\end{equation}
which for sufficiently slowly-varying fields can even take on
values $<-1$.

\subsection{Fluctuation formalism}

Using the synchronous perturbation formalism first developed by
Lifshitz and Khalatnikov, the study of scalar perturbations, $h$,
around the FRW metric can be reduced to investigate the solutions
of the equation for the $k$th mode [29]
\begin{equation}
\mu''+ \left(k^2 c_s^2-\frac{a''}{a}\right)\mu = 0 ,
\end{equation}
where $'=d/d\eta$, with $\eta=\int dt/a$ being the conformal time,
$c_s^2$= const. and the function $\mu$ is defined as
\begin{equation}
\mu=\frac{(h'+\alpha\gamma h)a}{\alpha\sqrt{\gamma}c_s}
=\frac{i\left(\dot{H}h-H\dot{h}\right)a}{H\sqrt{\dot{H}}c_s} ,
\end{equation}
with $^.=d/dt$, $\alpha=a'/a=\dot{a}$,
$\gamma=1-\alpha'/\alpha^2=1-\ddot{a}/\dot{a}^2$, the scale factor
$a(t)$ being given by Eq. (2.8) and
\begin{equation}
a(\eta)=\left[\frac{(1+3\omega)\eta}{2}\right]^{2/(1+3\omega)} ,
\end{equation}
where
\begin{equation}
\eta=\frac{2}{1+3\omega}a(t)^{(1+3\omega)/2} .
\end{equation}

At first sight it seems that the small-scale fluctuations must
grow when $c_s^2<0$. In fact, taking the limit for large $k$ in
Eq. (3.8) we obtain that the solution should be a growing mode for
$c_s^2<0$. But this is a growing mode in the conformal time
$\eta$, not in the physical time $t$, with $\eta$ and $t$ being
related to each other by means of Eq. (3.11). It can be readily
seen by using Eq. (2.8) that a growing mode in $\eta$ necessarily
implies a decreasing mode in $t$ for any $\omega<-1/3$
characterizing an accelerating universe. We note, moreover, that
the function $\mu$ is pure imaginary for $c_s$ real and real for
pure imaginary $c_s$. In the present model, we have from Eq.
(2.11) that $c_s^2=\omega$, Thus, if $c_s^2=\omega$ is assumed to
be constant, all the problem reduces to study a simple
differential equation given by
\begin{equation}
\mu''+\left[k^2c_s^2 -
\frac{2(1-3c_s^2)}{(1+3c_s^2)^2\eta^2}\right]\mu=0 .
\end{equation}
If the range of equation of state parameter would be extended to
include the regime $0<\omega<1/3$, then in that regime all would
happen like in the usual case, that is the perturbations would
oscillate for $k^2c_s^2>>2(1-3\omega)/[(1+3\omega)^2\eta^2]$ and
had an exponential behaviour for
$k^2c_s^2<<2(1-3\omega)/[(1+3\omega)^2\eta^2]$. However, if we
extend the range of $\omega$-values to also cover values
$\omega>1/3$, then gravity would behave like a repulsive force and
played on the same team as pressure does. For the most interesting
range where the universe shows an accelerating expansion,
$\omega<-1/3$, however, it is the pressure what plays on the same
team as gravity does. In that case, if
$k^2c_s^2>>2(1-3\omega)/[(1+3\omega)^2\eta^2]$, then the solution
to the differential perturbation equation reads
\begin{equation}
\mu\propto e^{\left(k|c_s^2|^{1/2}\eta\right)}=\exp\left(\pm
k|c_s^2|^{1/2}\frac{2T^{\frac{1+3\omega}{3(1+
\omega)}}}{1+3\omega}\right) ,
\end{equation}
with
\[T=a_0^{3(1+\omega)/2}+\frac{3}{2}(1+\omega)t .\]

In the range of state equation parameters $-1/3>\omega=c_s^2>-1$
it can be seen that $\mu$ decreases with time $t$, from an initial
value
\begin{equation}
\mu=\mu_0\propto\exp\left[\frac{\pm 2k|c_s^2|^{1/2}}{1
+3c_s^2}a_0^{(1+3c_s^2)/2}\right] ,
\end{equation}
at $t=0$, to become a minimum $\mu\propto 1$, as
$t\rightarrow\infty$. Also for the case where $\omega=c_s^2 <-1$
$\mu$ starts with a value given by Eq. (3.14) and becomes $\propto
1$ as one approaches the big rip time $t_*$. That decrease of
density fluctuations with small wavelengths clearly separates from
what was previously predicted by Carturan and Finelli, and Sandvik
et al. [20], being also in sharp contrast with what arises from
the equation that describes the evolution of small perturbations
in Newtonian physics, that is [30],
\begin{equation}
\ddot{\epsilon}_1 +\left(k^2 c_{\ell}^2-4\pi
G\epsilon_0\right)\epsilon_1 =0 ,
\end{equation}
where $c_{\ell}^2=p_1/\epsilon_1$, with $p_1$ and $\epsilon_1$
being the pressure and energy density first order perturbations.
In fact, for $c_{\ell}^2<0$, it is obtained that $\epsilon_1$
increases exponentially with time so that
\begin{equation}
\epsilon_1\propto \exp\left(\pm \sqrt{k^2|c_{\ell}^2|+ 4\pi
G\epsilon_0}t\right) .
\end{equation}

Although the mathematical reason for having obtained so a
different behaviour in our estimate for the time evolution of
density perturbations must be addressed to the specific form with
which the scale factor used in this paper depends on $\omega$
(which generalizes more restricted possible expressions), the
physical reason why the short-scale fluctuations are stable in our
model, even in the case where no contribution from baryons and
dark matter is explicitly considered, can be ultimately
interpreted to be the unified character of that model where dark
matter is also implicitly present.

We shall turn next to derive general analytical solutions for the
density contrast, $\delta$, in our tachyon model, thereby checking
and refining the conclusions attained so far in the above
heuristic model. That study will allow us, moreover, to deal in a
proper way with the important issue of the initial conditions for
fluctuations. If we assume that the tachyon field is given by Eq.
(2.9), the tachyon field theory considered in Sec. II is fully
equivalent to the simple theory of a perfect fluid with constant
equation of state $P=\omega\rho$ and squared sound speed
$c_s^2=\omega$ in a flat universe with a scale factor given by
either Eq. (2.8) if $\omega>-1$ or Eq. (2.13) if $\omega<-1$.
Thus, even though the density fluctuations of a perfect fluid
generally behaves differently from the density fluctuations of a
scalar field, in an adiabatic approach we can use the relativistic
analog of the Newtonian first order perturbation equation in order
to describe fluctuations of the density contrast. In our present
case, from the form of the energy density and the scale factor one
can derive an expression for the relativistic analog of the
Newtonian first order perturbation equation in Fourier space [31]
which reads for a density contrast perturbation $\delta_k$ with
wave vector $k$
\begin{equation}
\delta_k '' +\frac{1}{2}\left(1-9c_s^2\right)\delta_k '
-\left[\frac{3}{2}\left(1+2c_s^2-3c_s^4\right)-
\left(\frac{kc_s}{Ha}\right)^2\right]\delta_k = 0 ,
\end{equation}
where now $'=d/d\ln a$ and we have used $\omega=c_s^2$. Due to the
presence of a first-order derivative term in Eq. (3.17), also in
this case the taking of the simple large $k$ limit in this
equation when $c_s^2<0$ cannot by itself uncover the real
behaviour of small-scale fluctuations with physical time $t$. An
explicit integration of this differential equation should also be
done. For this to be accomplished, one can now re-write Eq. (3.17)
in the form
\begin{eqnarray}
&&a^2\delta_k '' +\frac{3}{2}a\left(1-3c_s^2\right)\delta_k '
\nonumber\\ && -\left[\frac{3}{2}\left(1+2c_s^2-3c_s^4\right) -k^2
c_s^2a^{1+3c_s^2}\right]\delta_k =0,
\end{eqnarray}
where now $'=d/da$ and we have used Eq. (2.8). Since $c_s^2<0$
this differential equation admits a general analytical solution
which can be expressed in terms of modified Bessel functions [32].
Depending on the initial conditions that solution can be taken to
be either some of the following ones or some of their combinations
\begin{widetext}
\begin{equation}
\delta_k = a^{-(1-9c_s^2)/4}e^{\frac{i\pi(5+
3c_s^2)}{4(1+3c_s^2)}}
I_{\frac{5+3c_s^2}{2(1+3c_s^2)}}\left(\pm\frac{2k|c_s^2|^{1/2}}{1+
3c_s^2}a^{(1+3c_s^2)/2}\right) ,
\end{equation}
\begin{equation}
\delta_k = \pm\frac{2}{i\pi}a^{-(1-9c_s^2)/4}e^{ \frac{i\pi(5+
3c_s^2)}{4(1+3c_s^2)}}
K_{-\frac{5+3c_s^2}{2(1+3c_s^2)}}\left(\pm\frac{2k|c_s^2|^{1/2}}{1+
3c_s^2}a^{(1+3c_s^2)/2}\right)
\end{equation}
for $-1/3>c_s^2>-5/3$,
\begin{equation}
\delta_k = \pm\frac{2}{i\pi}a^{-(1-9c_s^2)/4}e^{- \frac{i\pi(5+
3c_s^2)}{4(1+3c_s^2)}}
K_{\frac{5+3c_s^2}{2(1+3c_s^2)}}\left(\pm\frac{2k|c_s^2|^{1/2}}{1+
3c_s^2}a^{(1+3c_s^2)/2}\right)
\end{equation}
for $c_s^2\leq -5/3$.
\end{widetext}

We must now set the initial conditions for these perturbations. If
we assume that dark energy is present in the universe from its
very origin at $a=0$, the simplest and most natural condition that
such perturbations should satisfy is that they ought to tend to
vanishing values as one approaches the initial singularity both
for $\omega>-1$ and $\omega<-1$. One should then set $\delta_k=0$
as $a\rightarrow 0$. In the limit $a\rightarrow 0$ (i.e. for
arbitrarily large values of the argument $z$ of the modified
Bessel function, ${\it B}(\pm z)$), we have for ${\it B}(\pm
z)=I(\pm z)$
\begin{equation}
\delta_k\propto a^{-(1-3c_s^2)/2}\exp\left(\pm
\frac{2k|c_s^2|}{1+3c_s^2}a^{(1+3c_s^2)/2}\right)
\end{equation}
which vanishes for the + sign (i.e. for $-\pi<{\rm
arg}z\leq\pi/2$) and blows up for the - sign (i.e. for $\pi/2 <
{\rm arg}z\leq\pi$), and for ${\it B}(\pm z)=K(\pm z)$,
\begin{equation}
\delta_k\propto a^{-(1-3c_s^2)/2}\exp\left(\mp
\frac{2k|c_s^2|}{1+3c_s^2}a^{(1+3c_s^2)/2}\right),
\end{equation}
which vanishes for the lower sign and blows up for the upper sign.
Then the above initial conditions imply that we have to choose the
modified Bessel functions $I(+z)$ and $K(-z)$, or some combination
of them, for universes which start being dominated by dark energy
at $a=a_0$. Note that at $a=a_0$, i.e. at the time when dark
energy starts to dominate, $\delta_k$ consistently takes on
finite, nonzero values. It follows that the universe is stable
under perturbations $\delta_k$. Moreover, if the initial value of
the perturbation at $a=a_0$ is small enough then the fluctuation
will first grow up to a given maximum value to be steadily damped
thereafter, vanishing as $a\rightarrow\infty$, both for
$\omega>-1$ and $\omega<-1$. There would then be an interval of
scale factor values where these fluctuations had observable
effects. Since the sound speed is pure imaginary in all the
considered cases, fluctuations will gravitationally collapse in an
accelerated fashion to form, together with observable matter, some
structures with cosmological interest.

We note finally that consistency of using Eq. (3.17) with the
above initial conditions is manifested, if not guaranteed, by the
feature that they lead to exactly the same conclusions as those
obtained from the synchronous model. A caveat is worth mentioning
however. Once we have discussed fluctuations in the linear phase,
it appears most convenient to consider the fluctuations in the
non-linear phase, as these could become eventually dominant in
some regimes. Such a study is left for a future work.

On the other hand, another caveat should also be mentioned. Even
though our solutions apply to scales in the sub-horizon regime, if
the universe is expanding in an accelerating fashion then it would
also be natural to consider that sooner or later all scales will
become larger than the horizon so that the Newtonian approximation
would break down, specially when the limit $t\rightarrow \infty$
is approached for $\omega>-1$. In the case that $k^2 c_s^2 <<
2(1-3\omega)/[(1+3\omega)^2 \eta^2]$, the solutions to, for
example, Eq. (3.12) can be approximated to
\[\mu=A\eta^{2/(1+3c_s^2)}+B\eta^{-(1-3c_s^2)/(1+3c_s^2)} ,\]
with $A$ and $B$ arbitrary constants. We then see that for
$c_s^2<-1/3$, $\mu$ will grow with time $t$.

\subsection{Fluctuations in the contracting phase}

Once we have checked that our tachyon model for an accelerating
universe is stable against density fluctuations both when
$\omega>-1$ and for $\omega<-1$ up to the big rip time $t_*$, we
next proceed to briefly consider what happens with such
fluctuations when $\omega<-1$ at times $t>t_*$. This study is of
some interest for the following reason. Even though the big rip
corresponds to a curvature singularity that does not strictly
allow for causal connections between the regions before and after
the big rip, this might still be circumvented by connections
between these regions rendered physically plausible by means of
wormholes inflated by the accelerated expansion [13] or phantom
energy accretion [33]. It can be seen however that if we restrict
ourselves to the case $\omega<-1$ in Eq. (3.13) then for $t>t_*$
there will be a huge growth of the density of perturbations with
small wavelengths, so rendering the evolution of the universe
after the big rip fully unstable. On the other hand, in order for
studying the evolution of the fluctuations after $t_*$ in some
more detail, one must introduce the condition that $\delta_k =0$
as $t\rightarrow t_*$ (i.e. as $a\rightarrow\infty$). Now, for
$a\rightarrow\infty$ (i.e. as $z$ tends to zero), we have
\begin{equation}
\delta_k[I(\pm z)]\propto a^{1+3c_s^2} \rightarrow 0 ,
\end{equation}
\begin{equation}
\delta_k[K(\pm z)]\propto a^{1+3c_s^2} \rightarrow 0 ,
\end{equation}
if $-1>c_s^2>-5/3$, and
\begin{equation}
\delta_k[K(\pm z)]\propto a^{-3(1-c_s^2)/2}\rightarrow 0 ,
\end{equation}
for $c_s^2\leq -5/3$. We have to choose then the modified Bessel
functions $I(-z)$ and $K(-z)$, or some given combination of them,
for universes starting to evolve just after the big rip
singularity. It can be then checked that such conditions also
predict that the contracting universe that starts evolving just
after the big rip is unstable to the considered fluctuations.

\section{Structure formation in a tachyon dark-energy universe}

In this section we will briefly consider the fate of the
fluctuations $\delta_k$ once they have formed and evolved to their
nearly largest size. From the onset of the epoch where dark energy
starts dominating the formed overdense regions in the universe
will follow an evolution which is governed by a stuff
characterized by negative pressure and speed of sound squared.
That will inexorably makes the force involved in such fluctuations
to be attractive as pressure will now contribute on the same
attractive pattern as gravity. Thus, the overdense regions would
immediately start to undergo an accelerated gravitational collapse
which involved both dark matter and dark energy, and hence any now
subdominant possible visible matter that could be more or less
associated with them, towards a central singularity with infinite
density. In practice, since the stuff making up tachyon dark
matter-energy is expected to be spinless and chargeless, in the
present case only some macroscopic dissipative processes, other
than pressure, spin degeneracy or electric charge repulsion, could
intervene to convert kinetic energy of collapse into random
motions before the singularity. Black holes mainly made from dark
matter could be thereby most easily formed at the end of the
collapse, though the residual dissipative forces could still stop
the collapse before reaching the final dynamical equilibrium.

\subsection{Accelerated spherical collapse}

We shall assume in what follows a spherical collapse model with no
shell-crossing, so that we can ignore the spatial dependence of
the involved fields [34]. In principle, the evolution of a
spherical overdense patch with scale radius $R(t)$ containing
independent dark matter (with density $\rho_{{\rm DM}}$) and dark
energy (with density $\rho=p/\omega$) is governed by the
Raychaudhuri equation [30]
\begin{equation}
3\ddot{R}=-4\pi GR\left[\rho_{{\rm DM}}+(1+3\omega)\rho\right] .
\end{equation}
For our unified model, the first term in squared brackets of this
equation becomes redundant because dark matter is already
contained in the unified tachyon dark description encapsulated in
the term containing density $\rho$. Thus, for the case being
considered, Eq. (4.1) reduces to:
\begin{equation}
3\ddot{R}=-4\pi GR(1+3\omega)\rho .
\end{equation}
Now, in order to check consistency of our unified dark energy
model, let us first apply Eq. (4.1) to an energy density as given
in Eq. (2.2) which, when expressed in terms of variable
$T=a_0^{3(1+\omega)/2}+3(1+\omega)t/2$, reads
\begin{equation}
\rho=\frac{3}{8\pi GT^2} .
\end{equation}
That is, one is considering the universe as a whole as a
perturbation and, therefore, one should recover the scale factor
(2.8) as a solution from Eq. (4.1). This is in fact the case, but
we obtain also another solution given by
\begin{equation}
R=a_c=T^{(1+3\omega)/[3(1+\omega)]}=a(t)^{(1+3\omega)/2} .
\end{equation}
Note that this additional solution will vanish as
$a\rightarrow\infty$ both for $\omega>-1$ and for $\omega<-1$ and
represents therefore a collapsing universe.

For a spherical overdense patch with scale radius $R(t)$ within
the universe, the Raychaudhuri equation will be
\begin{equation}
3\ddot{R}=-4\pi G R(1+3\omega)(\rho+\mu) ,
\end{equation}
where $\rho$ is given by Eq. (4.3) and we have restricted
ourselves to consider perturbations of the energy density that
correspond only to small wavelengths; that is perturbations that
can be described by
\begin{eqnarray}
&&\mu=\mu_0\exp\left[\frac{-2k|c_s^2|^{1/2}
T^{(1+3\omega)/[3(1+\omega)]}}{1+3\omega}\right]\nonumber\\
&&=\mu_0\exp\left(\frac{-k|c_s^2|^{1/2}a_c}{1+3\omega}\right) .
\end{eqnarray}
Obtaining an exact analytical solution to the differential
equation (4.7) is very difficult and we will simply consider in
what follows approximate asymptotic solutions when $T$ is very
large or very small, for particular values of $\omega$. On the
regime of $\omega>-1$ and $T$ large, we take the observationally
plausible value $\omega=-13/15$ so that the differential equation
(4.7) can then be approximated to
\begin{equation}
R''-\frac{128\pi G}{153}\left(\mu_0 +\frac{3}{8\pi
GT^2}\right)R\simeq 0 .
\end{equation}
A solution is then again given in terms of the modified Bessel
function $\wp$ [32]
\begin{equation}
R\simeq T^{1/2}\wp_{\sqrt{53/180}}\left(\sqrt{\frac{128\pi
G\mu_0}{135}} T\right).
\end{equation}
For small values of $T$ on the phantom regime at the particular
case that $\omega=-5/3$, the approximate differential equation
(4.7) will become
\begin{equation}
R''-\frac{64\pi G}{27}\left(\mu_0+\frac{3}{8\pi
GT^2}\right)R\simeq 0 ,
\end{equation}
which again admits an approximate solution in terms of modified
Bessel functions $\wp$, that is [32]
\begin{equation}
R\simeq T^{1/2}\wp_{\sqrt{41/36}}\left(\sqrt{\frac{64\pi
G\mu_0}{27}} T\right).
\end{equation}

Choosing now for $a_0$ a sufficiently large value then $T$ will
always be large enough for $\omega>-1$ and small enough for
$\omega<-1$. Since the most natural and sufficient initial
condition in the present case is that $R<<a$ initially, then the
only solutions which satisfy the above initial condition are for
$\omega=-13/15$ (and generally for any $\omega>-1$) the one with
$\wp=K$, i.e.
\begin{equation}
R\propto\exp\left(-\sqrt{\frac{128\pi G\mu_0}{135}}T\right) ,
\end{equation}
and for $\omega=-5/3$ (and generally for any $\omega<-1$) the one
with $\wp=I$, i.e.
\begin{equation}
R\propto T^{\frac{1}{2}\left(1+\sqrt{\frac{41}{9}}\right)} .
\end{equation}
Now these two solutions tend to vanish as one lets
$a(t)\rightarrow\infty$, so indicating a complete accelerated
gravitational collapse of the overdense patches, which can be
modulated by the residual dissipative forces. Although the
possibility that the black holes that eventually formed as the
final state of such collapses actually be those supermassive black
holes ultimately driving formation of the observed galaxies
[23,35] is just a conjecture we want to advance here (see Sec. V),
it already appears interesting to consider that possibility as a
quite promising way through which the studied fluctuations may
show up observationally.

\subsection{Loss of dark energy}

On the other hand, dark energy does not necessarily follow the
collapse of dark matter inside the overdensity regions and
therefore there could be an energy loss of dark energy $\Gamma$ in
these regions. In the present case this energy loss can be
described by the equation [36]
\begin{equation}
\Gamma=\dot{\phi}\ddot{\phi}+
\left(1-\dot{\phi}^2\right)\left(3\frac{\dot{R}}{R}\dot{\phi}^2
+\frac{\dot{V}}{V}\right) ,
\end{equation}
with $R$ the scale radius of an average overdense region, and
$\phi$ and $V=V(\phi)$ the scalar field and the potential on that
region. There are important technical difficulties which would
appear if we try to calculate $\Gamma$ using the expressions of
$R$ discussed in the precedent subsection. In order to evaluate on
which cases $\Gamma\neq 0$ we tentatively shall simply interpret
for a moment the collapsing solution
$a_c=T^{(1+3\omega)/[3(1+\omega)]}$ to the Raychaudhuri equation
like though if it represented the scale radius of an overdensity
patch, i.e. $R=a_c$, and then use the two ans\"{a}tze considered in
Sec. IV for the partition of the full energy density and pressure
of the unified model into dark matter and dark energy components.
In such a toy description we have for the not partitioned theory
\begin{equation}
H=\frac{\dot{R}}{R}=\frac{1+3\omega}{2T}
\end{equation}
\begin{equation}
\dot{\phi}=\sqrt{\frac{2(1+\omega)}{1+3\omega}}
\end{equation}
\begin{equation}
V=\frac{3(1+3\omega)^{3/2}(\omega-1)^{1/2}}{32\pi GT^2} .
\end{equation}
Since no partition of energy density and pressure has been made
yet, the insertion of Eqs. (4.16)-(4.18) into Eq. (4.15)
consistently leads to the result that there is no loss of any
energy on the overdense region and $\Gamma=0$. The use of the
first ansatz for partitioning dark energy and pressure given by
Eq. (3.4) leads to $\omega_v=-1$, $\dot{\phi}_v=0$, $H=-T^{-1}$
and $V=3/(4\pi GT^2)$, and hence again $\Gamma=0$, as it was also
expected for a cosmological constant. However, using the partition
given by Eq. (3.6), we obtain $\omega_v=1-\dot{\phi}^{-2}$, and
finally
\begin{equation}
\Gamma=\frac{3\omega}{2T}\left(\frac{2\omega^2
+3\omega-1}{1-\omega}\right) ,
\end{equation}
which shows a loss of dark energy in the overdense region for any
$\omega>(3+\sqrt{17})/4$. We then notice that the collapse
tachyonic overdense patches will mainly affect the dark matter
component only if the dark energy component does not correspond to
a positive cosmological constant with $\omega_v=-1$. For in that
case the collapse would equally affect both components.

\section{A conjecture on galaxy formation}

Having shown that the evolution of density perturbations in the
unified tachyon theory does not lead to a catastrophic growth of
perturbations with small wavelengths, but to a
negative-pressure-assisted super accelerated gravitational
collapse of the dark matter and energy involved in the overdense
regions, leaving rapidly formed black holes which may have a large
mass, one should now consider the observational consequences from
the formation of such holes. At first sight, the results obtained
in the present paper could in fact be disregarded under the claim
that no giant black holes of the kind discussed in it have been
hitherto observed. We are going now to conjecture nevertheless
that it is precisely a family of rapidly forming huge black holes
with masses analogous to those of the black holes resulting from
the accelerated collapse of overdense regions in tachyon unified
dark energy discussed in this paper which have been discovered in
recent years at the center of most - possibly all - quasars (QSOs)
and current luminous galaxies [23]. The rationale supporting such
a conjecture runs up along the following points.

\noindent {\bf 1.} The idea that bright galaxies may harbor
supermassive black holes (SMBHs) actually dates back to the year
1969 and was first advanced by Lynden-Bell [37]. Today it is a
rather common belief that most - if not all - galaxies have a
central SMBH in their bulge [23]. It is also currently thought
that at the time of the formation of one galaxy its present
central black hole started also to be formed [36].

\noindent {\bf 2.} The way in which SMBHs were formed is still
unclear. They might have a primordial origin, be formed from
population III stars and their resulting merging, or otherwise
[38]. The physical process involved could respectively be either
by directly collapsing from external pressure in the first instant
of big bang or by slow accreation of matter starting from a
stellar size, or otherwise. The idea that SMBHs may have passed
through an intermediate observable mass stage along a very long
accreating process is now gaining increasing support [39].

\noindent {\bf 3.} The epoch of galaxy formation is thought to be
very broad, extending from $z\sim 20$ to $z\sim 2$. The galaxies
were formed from perturbations created at $z\sim 1400$. However,
galaxies with redshifts $\geq 2-3$ are not directly observable.
Since a baryonic gas inside a dark matter potential collapses and
forms stars, any resulting massive stars rapidly evolving and
feedbacking heavy elements into the gas, the production of first
heavy elements can be approximately equated to the epoch of bright
galaxy formation. Metal lines have been so far identified in QSO
spectra [40] indicating that, prior to $z=2.5-3$, the heavy metal
line strengths are fairly low, with the largest strengths
appearing within the interval $z=1.5-2.5$. Thus, non luminous
galaxies could have existed from nearly the primordial epoch, but
the formation of a sufficiently high density of stars in them
making these galaxies bright appears to be fairly recent.

\noindent {\bf 4.} The observational discovery that there exists a
simple relation between the mass of central SMBHs and the speed or
the sigma of the stars in the surrounding galaxy bulges [41],
which was anticipated by the Silk-Rees theory [42], seems then to
indicate that most of such central SMBHs reached -or were created
with - their present huge maturity mass just at around $z\sim
2-3$, nearly at, or shortly before, the epoch when the giant black
holes discussed in the present paper probably started to be
collapsed in an accelerated way. The onset of that epoch should
nevertheless be marked by the farthest supernova at $z=1.7$ [43],
that is to say around $z\sim 1$. Moreover, there exist QSOs at
$z\sim 6$ and bright galaxies at $z\sim 3$ which must harbor a
SMBH at their central bulge. At first sight, these features would
prevent SMBHs formed by the collapse of overdense tachyon dark
energy regions to be suitable candidates for being the central
holes that would induce bulge star creation and hence bright
galaxy formation. Notwithstanding, there are at least two main
arguments against that conclusion. On the one hand, it has been
shown [44] that in the presence of a cosmological constant the
galaxy formation epoch must have started much later and lasted
much longer, possibly all the way to the present. Similarly,
taking into account the Friedmann equation for flat geometry, if
instead of the cosmological constant density $\lambda_0
=\Lambda/(3H_0^2)$ of Ref. [44], we introduce the dynamical
dark-energy density considered in this paper, then a theoretical
onset for bright galaxy formation can be derived by replacing
$\lambda_0$ in the formula provided in that reference for the
quantity $8\pi G\rho/(3H_0^2)$. Using then Eqs. (2.2), (2.3) and
(2.10) one can finally have
\begin{equation}
1+z_{{\rm GF}}=\left[H_0^2 T^2 \Omega_0\sinh^2\left(\frac{3t_{{\rm
GF}}}{2T}\right)\right]^{-1/3} ,
\end{equation}
where $H_0$ is the current value of the Hubble constant and
$t_{{\rm GF}}$ is the galaxy formation time, $t_{{\rm GF}}=t_d
+t_c$, with $t_d$ the time needed for a galaxy to start forming
stars once the region with local Hubble parameter $H'$ has
collapsed, and $t_c$ is the collapse time which is in the present
case determined by
\begin{equation}
t_c=
2\int_0^{x_{max}}\left(U+\frac{2}{x}+\frac{x^2}{T^2}\right)^{-1/2}
dx ,
\end{equation}
where if the mass of the overdense region of scale radius $r$ is
denoted by $m$,
\begin{equation}
x=r(Gm)^{-1/3}
\end{equation}
\begin{equation}
U=4^{1/3}\left[\left(\frac{H'}{H}\right)^2
-\Omega(1+\bar{\delta})-\frac{1}{H^2
T^2}\right]\left(\frac{1+\bar{\delta}}{\Omega H^2}\right)^{-2/3}
\end{equation}
and
\begin{equation}
U+\frac{2}{x_{max}}+\frac{x_{max}}{T^2}=0 ,
\end{equation}
with $\Omega=\bar{\rho}/(3H^2)$, $\bar{\rho}$ and $H$ respectively
being the mean matter density and Hubble parameter of the
universe. We can check that also in this case for suitable initial
values of the observable density contrast $\bar{\delta}_i$, the
galaxy formation epoch is shifted toward later times by some two
$z$-units if currently $8\pi
G\rho/(3H_0^2)\simeq\Lambda/(3H_0^2)\simeq 0.8$, while it lasts
quite longer, possibly all the way to the present, too, such as it
occurred in the presence of a positive cosmological constant [44],
except in that in the present theory $z_{{\rm GF}}$ depend on time
$t$ through the parameter $T$. It is worth noticing e.g. that in
the phantom regime where $\omega<-1$, as one approaches the big
rip at $T=0$, $t_c\rightarrow 0$ and $z_{GF}\rightarrow 0$, so
that, consistent with the ripping off of all materials in the
universe, observers would detect no galaxies at all.

On the other hand, and even more importantly, if a direct
interaction term between the two components $\rho_v$ and
$\rho_{DM}$ transferring energy from one another is assumed to
exist and have the form
\begin{equation}
\gamma= -3\ell_P\sqrt{\Omega_v}\omega\rho_{DM} ,\;\;
\Omega_v=\frac{\rho_v}{3H^2} ,\;\; \ell_P^2=\frac{8\pi G}{3} ,
\end{equation}
then it can be readily seen that the theory discussed in this
paper becomes completely equivalent to the model of coupled dark
matter and energy developed by Amendola [45]. Therefore one should
expect that for our model the accelerating expansion would start
at a maximum redshif of around $z\sim 5$ [45] in a way which is
consistent with the type Ia supernovae Hubble diagram, including
the farthest known supernova at $z=1.7$, while allowing for
structure formation during the acceleration regime. In fact, if
there was an epoch at $z_{acc}$ in the past before which uncoupled
baryons with critical density $\Omega_b$ dominated a decelerating
flat universe filled also with tachyon fields, then in the present
theory with $\gamma$ given by Eq. (5.6) that epoch would be given
by
\begin{equation}
1+z_{acc}=\left[-(1+3\omega)\left(\frac{1-
\Omega_b}{\Omega_b}\right)\right]^{1/(3\omega)} ,
\end{equation}
instead of the currently assumed value [45]
\begin{equation}
z_{acc}=\left[-(1+ 3\omega)\left(\frac{\Omega_{\phi}}{\Omega_{DM}
+\Omega_b }\right)\right]^{1/(3\omega)}\sim 1 .
\end{equation}
Finally, according to Eq. (5.7) for the allowed values of
$\Omega_b$, on the phantom regime ($\omega<-1$) the larger
$|\omega|$ the smaller $z_{acc}$.

\noindent {\bf 5.} Thus, since the nature and properties of
nonrotating, chargeless black holes of any sizes do not
qualitatively distinguish the original characteristics of the
stuff out of which they were formed, the difficulties encountered
so far to justify presently considered mechanisms describing how
central SMBHs were formed within galaxy bulges [46] might all be
circumvented if we conjectured that, since it turns out that the
tachyon field perturbations in density contrast took place with a
spatial distribution nearly matching that of matter fluctuations,
and they initially may rapidly have grown during a more or less
long period, shortly after the onset of tachyon dark-energy
domination at a sufficiently large $z$, giant black holes mainly
made up of dark matter and energy, formed by the kind of quickly
accelerated collapsing procedure considered in Sec. IV, started
forming at the galactic centers and very rapidly reached a state
where they were able to switch on stars in the surrounding bulge
by means of a process similar to the Silk-Rees mechanism [42].
That conjecture would seem to decide the debate on which came
first, the black hole or the galaxy, in favor of the latter, but
required the existence of galaxies devoid of high star densities
until their central SMBHs were formed.

As to the influence that ordinary matter sources may have on the
stability of tachyon dark energy and its accelerated collapse, we
note that if these sources contribute as a perturbation, one can
analyze the whole model via phase space [47]. Then the potentials
(2.10) and (2.16) and the solutions (2.8) and (2.13) would
correspond to dynamical attractors for, respectively, tachyon dark
and phantom energy domination with their respective cosmological
effects. For these attractors we have [47]
$\Theta=VV''/(V')^2=3/2$, at critical points defined by $8\pi
G\lambda_c$ which equals $3\sqrt{1+\omega}/T$ for dark energy and
$-3\sqrt{|\omega|-1}/T$ for phantom energy. It follows then that:
(i) the models of Secs. II-IV are stable to the presence of small
proportions of ordinary matter, and (ii) even though that ordinary
matter would roughly follow distribution of the dark matter and
energy, one would expect it not to significantly undergo initial
accelerated collapse as it had to obey a different equation of
state. Thus, ordinary baryonic matter would essentially remain as
the stuff out of which stars are going to be formed.

\section{Conclusions and further comments}

The assumed dark energy content of the universe possibly reflects
the greatest unsolved problem of all of physics. This paper has
explored some potentially important observational consequences
from one of the most promising models suggested to describe dark
energy, the so-called tachyon model. Thus, by using tachyon-like
theories whose starting Lagrangians generally are inspired by
string theories, we have investigated the properties of general
FRW cosmological solutions that are fully consistent with the
whole dynamical structure of the theories. Assuming a general
equation of state $p=\omega\rho$, such solutions describe
accelerating universes in the interval $-1/3>\omega>-1$ and, among
other weird properties characterizing the phantom regime, all show
a big rip singularity in finite time when $\omega<-1$. It has been
generally shown as well that the entropy of a universe filled with
phantom energy is always positive, though its temperature becomes
definite negative. This keeps still phantom cosmology as a real,
observationally not excluded possibility.

Fluctuations taking place in both of such regimes have been
studied by using a synchronous perturbation formalism and the
relativistic analog of the Newtonian first order perturbation
theory. In both formalisms the result has been obtained that the
considered tachyon theory is stable to fluctuations along the
entire considered interval of the state equation parameter
$-5/3<\omega=p/\rho <-1/3$.

A potential problem with the tachyon models considered in this
paper stems from the imaginary value of the sound speed. In fact a
value $c_s^2<0$ implies occurrence of instability on scales below
the Jeans limit for scalar field fluctuations [20] which may grow
therefore exponentially. Whereas when these models are considered
as pure cosmic vacuum components this could in fact be regarded as
an actual problem, if the tachyon models are viewed as the sum of
two components [17], one describing the negative pressure vacuum
stuff and the other describing the dust-like cold dark matter
contributing $\Omega_m\sim 0.35$ and clustering gravitationally at
small scales, the results obtained in this paper for a negative
value for $c_s^2$ might instead be regarded as originating an
accelerated process of gravitational collapse inexorably leading
to the rapid formation of black holes in a way that may explain
some recent observations in galaxies and superclusters that
concern gravitationally collapsed objects such as supermassive
black holes. Alcaniz and Lima suggested nevertheless that a
quintessence component pushes back to higher redshift the epoch of
formation of some galaxies [48]. The analysis carried out by these
authors is based however on a simple dark energy component which
is uncoupled to dark matter. If one assumes that dark matter and
energy are coupled to each other, then since it turns out that the
perturbations in density contrast may initially grow during a
period, we have in fact advanced the conjecture that supermassive
black holes with masses ranging from $10^6$ to $10^9$ solar masses
which have been observed at the center of most galactic bulges
could actually be originated by the rapid accelerated collapse of
the gravitationally enlarged overdense regions made up of tachyon
dark matter and energy. Rather than being destructive influences
on their surroundings, the resulting giant black holes would then
have a very creative impact in the formation of the host galaxies.
This would offer a novel alternate resolution to the problem of
supermassive black hole formation. Much calculation and
observational scrutiny is however needed before such a conjecture
can be seriously considered. In particular, it appears that future
observations on supernovae of type Ia at high redshift will be
decisive in order to check acceptability of our conjecture.

\acknowledgements

\noindent The author thanks Carmen L. Sig\"{u}enza for useful
discussions and Yun-Song Piao and Fabio Finelli, J.S. Alcaniz and
J.A.S. Lima for valuable correspondence. This work was supported
by DGICYT under Research Project BMF2002-03758.


\begin{references}

\bibitem {1} S. Perlmutter {\it et al.},
Astrophys. J. 483, 565 (1997); S. Perlmutter {\it et al.} (The
Supernova Cosmology Project), Nature 391, 51(1998); P.M. Garnavich
{\it et al.} Astrophys. J. Lett. 493, L53 (1998); B.P. Schmidt,
Astrophys. J. 507, 46 (1998); A.G. Riess {\it et al.} Astron. J.
116, 1009 (1998).
\bibitem {2} M.S. Turner, {\it Dark energy and the new cosmology},
astro-ph/0108103, Contribution to the SNAP Yellow Book (Snowmass
2001); M. Turner and A.G. Riess, Astrophys. J. 569, 18 (2002).
\bibitem {3} B. Ratra and P.J.E. Peebles, Rev. Mod. Phys. 75, 559
(2002).
\bibitem {4} C. Wetterich, Nucl. Phys. B302, 645 (1988); B. Ratra
and P.J.E. Peebles, Astrophys. J. 325, L17 (1988).
\bibitem {5} C. Armend\'{a}riz-Pic\'{o}n, T. Damour and V. Mukhanov, Phys.
Lett. B458 (1999) 209; J. Garriga and V. Mukhanov, Phys. Lett.
B458 (1999) 219; T. Chiba, T. Okabe and M. Yamaguchi, Phys. Rev.
D62 (2000) 023511; C. Amend\'{a}riz-Pic\'{o}n, V. Mukhanov and P.J.
Steinhardt, Phys. Rev. Lett. 85 (2000) 4438; C. Amend\'{a}riz-Pic\'{o}n,
V. Mukhanov and P.J. Steinhardt, Phys. Rev. D63 (2001) 103510;
L.P. Chimento and A. Feinstein, Mod. Phys. Lett. A19, 761 (2004).
\bibitem {6} J.C. Jackson and M. Dodgson, Mon. Not. R. Astron. Soc.
297, 923 (1998); J.C. Jackson, Mon. Not. R. Astron. Soc. 296, 619
(1998); R.R. Caldwell, R. Dave and P.J. Steinhardt, Phys. Rev.
Lett. 80, 1582 (1998); L. Wang and P.J. Steinhardt, Astrophys. J.
508, 483 (1998); R.R. Caldwell and P.J. Steinhardt, Phys. Rev.
D57, 6057 (1998); G. Huey, L. Wang, R. Dave, R.R. Caldwell and
P.J. Steinhardt, Phys. Rev. D59, 063005 (1999); P.F.
Gonz\'{a}lez-D\'{\i}az, Phys. Rev. D62, 023513 (2000).
\bibitem {7} P.J. Steinhardt, L. Wang and I. Zlatev, Phys. Rev. Lett. 82, 896
(1999); ibid, Phys. Rev.D59, 123504 (1999); I. Zlatev and P.J.
Steinhardt, Phys. Lett. B459, 570 (1999).
\bibitem {8} A. Kamenshchik, U. Moschella, V. Pasquier, Phys.
Lett. B511, 265 (2001); N. Bilic, G.B. Tupper and R. Viollier,
Phys. Lett. B535, 17 (2001); M.C. Bento, O. Bertolami and A.A.
Sen, Phys. Rev. D66, 043507
\bibitem {9} A. Sen, JHEP 0204, 048 (2002); A. Sen, Mod. Phys.
Lett. A17, 1797 (2002); G.W. Gibbons, Phys. Lett. B537, 1 (2002) .
\bibitem {10} A.C. Baccigalupi, A. Balbi, S. Matarrase, F. Perrotta
and N. Vittorio, Phys. Rev. D65, 063520 (2002); M. Melchiorri, L.
Mersini, C.J. Odman and M. Tradden, Phys. Rev. D68, 043509 (2003);
M. Doupis, A. Riazuelo, Y. Zolnierowski and A. Blanchard, Astron.
Astrophys. 405, 409 (2003); L. Tonry {\it el al.}, Astrophys. J.
594, 1 (2003); J.S. Alcaniz, Phys. Rev. D69, 083521 (2004).
\bibitem {11}R.R. Caldwell, Phys. Lett. B545, 23 (2002).
\bibitem {12} B. McInnes, JHEP 0208, 029 (2002);
G.W. Gibbons, hep-th/0302199; A.E. Schulz and M.J. White, Phys.
Rev. D64, 043514 (2001); J.G. Hao and X. Z. Li, Phys. Rev. D67,
107303 (2003); S. Nojiri and S.D. Odintsov, Phys. Lett.B562, 147
(2003); B565, 1 (2003); B571, 1 (2003); P.Singh, M. Sami and N.
Dadhich, Phys. Rev. D68, 023522 (2003); J.G. Hao and X.Z. Li,
Phys. Rev. D68, 043501; 083514 (2003); X.Z. Li and J.G. Hao, Phys.
Rev. D69, 107303 (2004); M.P. Dabrowski, T. Stachowiak and M.
Szydlowski, Phys. Rev. D68, 067301 (2003); M. Szydlowski, W. zaja
and A. Krawiec, astro-ph/0401293; E. Elizalde and J. Quiroga H.,
Mod. Phys. Lett. A19, 29 (2004); V.B. Johri, astro-ph/0311293;
L.P. Chimento and R. Lazkoz, Phys. Rev. Lett. 91, 211301 (2003);
H.Q. Lu, hep-th/0312082; M. Sami, A. Toporensky, Mod. Phys. Lett.
A19, 1509 (2004); R. Naboulsi, gr-qc/0303007, Class. Quan. Grav
(in press); J.M. Cline, S. Jeon and G.D. Morre, hep-ph/0311312,
Phys. Rev. D (in press); X.H. Meng and P. Wang, hep-ph/0311070; H.
Stefancic, Phys. Lett. B586, 5 (2004); astro-ph/0312484, Eur.
Phys. J. C (in press); D.J. Liu and X.Z. Li, Phys. Rev. D68,
067301 (2003); A. Yurov, astro-ph/0305019; Y.S. Piao and E. Zhou,
Phys. Rev. D68, 083515 (2003);Y.S. Piao and Y.Z. Zhang,
astro-ph/0401231; H.Q. Lu, hep-th/0312082; L.P. Chimento and R.
Lazkoz, astro-ph/0405518; J.M. Aguirregabiria, L.P. Chimento, R.
Lazkoz, astro-ph/0403157; J.D. Barrow, Class. Quant. Grav. 21, L79
(2004); S. Nojiri and S.D. Odintsov, hep-th/0405078; E. Elizalde,
S. Nojiri and S.D. Odintsov, hep-th/0405034.
\bibitem {13} P.F. Gonz\'{a}lez-D\'{\i}az, Phys. Rev. D68, 084016 (2003).
\bibitem {14}  R.R. Caldwell, M. Kamionkowski and N.N. Weinberg,
Phys. Rev. Lett. 91, 071301 (2003).
\bibitem {15} P.F. Gonz\'{a}lez-D\'{\i}az, Phys. Lett. B586, 1 (2004).
\bibitem {16} P.F. Gonz\'{a}lez-D\'{\i}az, Phys. Rev. D68, 021303(R) (2003);
M. Bouhmadi and J.A. Jim\'{e}nez-Madrid, astro-ph/0404540.
\bibitem {17} T. Padmanabhan, Phys. Rev. D66, 021301 (2002);
T. Padmanabhan and T.R. Choudhury, Phys. Rev. D66, 081301 (2002);
J.S. Bagla, H.K. Jassal, T. Padmanabhan, Phys. Rev. D67, 063504
(2003)
\bibitem {18} P.F. Gonz\'{a}lez-D\'{\i}az, Phys. Rev. D69, 103512 (2004).
\bibitem {19} L.R.W. Abramo and F. Finelli, Phys. Lett. B575, 165
(2003).
\bibitem {20} D. Carturan and F. Finelli, Phys. Rev. D68, 103501
(2003); H.B. Sandvik, M. Tegmark, M. Zaldarriaga and I. Waga,
astro-ph/0212114; A. Frolov, L. Kofman and A. Starobinsky, Phys.
Lett. B545, 8 (2002).
\bibitem {21} L.R.W. Abramo, F. Finelli and T.S. Pereira,
astro-ph/0405041
\bibitem {22} L. Amendola, F. Finelli, C. Burigana and D. Carturan,
JCAP 0307, 005 (2003).
\bibitem {23} M.J. Rees, in: {\it Black Holes and Relativistic Stars},
edited by R.M. Wald (Chicago University Press, Chicago, USA,
1988); M.J. Rees, in: {\it Black Holes in Binaries and Galactic
Nuclei}, edited by L. Kaper, E.P.J. van den Heurel and P.A. Would
(Springer-Verlag, New York, USA, 2001); F.D. Machetto, in: {\it
Towards a New Millennium in Galaxy Morphology}, edited by D.L.
Block {\it et al.} (Kluwer, Dordrecht, USA, 1999).
\bibitem {24} P.T.P. Viana and A.R. Liddle, Phys. Rev. D57, 674
(1998); W. Hu, Astrophys. J. 506, 495 (1998); M. White, Astrophys.
J. 506, 485 (1998).
\bibitem {25} P.F. Gonz\'{a}lez-D\'{\i}az, Phys. Rev. D69, 063522 (2004).
\bibitem {26} J.A.S. Lima and J.S. Alcaniz, astro-ph/0402265 .
\bibitem {27} D. Youm, Phys. Lett. B531, 276 (2002); E. Verlinde,
0008140; J.L. Cardy, Nucl. Phys. B270, 186 (1986).
\bibitem {28} J.B. Hartle and S.W. Hawking, Phys. Rev. D13, 2188
(1976).
\bibitem {29} E.M. Lifshitz and I. Khalatnikov, Adv. Phys. 12, 185
(1963); L.P. Grishchuk, Phys. Rev. D50, 7154 (1994); J.C. Fabris
and J. Martin, Phys. Rev. D55, 5205 (1997).
\bibitem {30} J.P.E. Peebles, {\it The Large Scale Structure of
the Universe} (Cambridge University Press, Cambridge, UK, 1980).
\bibitem {31} T. Padmanabhan, {\it Structure Formation in the
Universe} (Cambridge University Press, Cambridge, UK, 1993).
\bibitem {32} M. Abramowitz and I.A. Stegun, {\it Handbook of
Mathematical Functions} (Dover, New York, USA, 1965).
\bibitem {33} P.F. Gonz\'{a}lez-D\'{\i}az, astro-ph/0404045, Phys. Rev.
Lett. (in press, 2004).
\bibitem {34} P. Fosalba and E. Gazta\~{n}aga, Mon. Not. R. Astron.
Soc. 31, 535 (1998) .
\bibitem {35} M.G. Haehnelt, astro-ph/0307378, in: {\it
Coevolution of Black Holes and Galaxy. Carnegie Observatories
Series}, Vol. 1, edited by L.C. Ho (Cambridge University Press,
Cambridge, UK, 2003).
\bibitem {36} C. van de Bruck, astro-ph/0401504 (to appear in
Astron. Astrophys.) .
\bibitem {37} D. Lynden-Bell, Nature 223, 690 (1969).
\bibitem {38} S.L. Shapiro, in: {\it Coevolution of Black Holes
and Galaxies. Carnegie Observatories Astrophysical Series}, Vol.
1, edited by L.C. Ho (Cambridge University Press, Cambridge, UK,
2003); C. Tyler, B. Janus and D. Santos-Noble, astro-ph/0309008
\bibitem {39} R.P. van der Marel, J. Gerssen, P. Guhathakurta,
R.C. Peterson and K. Gebhardt, astro-ph/0209314, Astron. J. (in
press); K. Gebhardt, R.M. Rich and L.C. Ho, Astrophys. J. 578, L41
(2002).
\bibitem {40} C.C. Steidel, M. Giavalisco, M. Pettini, M.
Dickinson and K.L. Adelberger, Astrophys. J. 462, L17 (1996) .
\bibitem {41} K. Bebhardt {\it et al.}, Astrophys. J. 539, L13
(2000); L. Farrarese and D. Merritt, Astrophys. J. 539, L9 (2000)
.
\bibitem {42} J. Silk and M.J. Rees, Astron. Astrophys. 331, 1
(1998); M.G. Haehnelt, P. Natarajan and M.J. Rees, Mon. Not. R.
Astron. Soc. 3000, 817 (1998) .
\bibitem {43} A.G. Riess {\it et al.}, Astrophys. J. 560, 49
(2001); A.G. Riess {\it et al.}, astro-ph/0402512 (Astrophys. J.,
to appear).
\bibitem {44} H. Martel, Astrophys. J. 421, L67 (1994).
\bibitem {45} L. Amendola, Mon. Not. R. Astron. Soc. 342, 221
(2003).
\bibitem {46} J. Silk, Space Sci. Rev. 100, 41 (2002).
\bibitem {47} J.G. Hao and X.Z. Li, astro-ph/0309746.
\bibitem {48} J.S. Alcaniz and J.A.S. Lima, Astrophys. J. 550,
L133 (2001).

\end{references}
\end{document}